# Incoherently pumped high-power linearly-polarized single-mode random fiber laser: experimental investigations and theoretical prospects

*Jiangming Xu,[1,2,*] Zhaokai LOU,[1] Jun Ye,[1] Jian Wu,[1,2] Jinyong Leng,[1,2] Hu Xiao,[1,2] Hanwei Zhang,[1,2] and Pu Zhou[1,2,**]*

[1]*College of Optoelectronic Science and Engineering, National University of Defense Technology, Changsha 410073, PR China*
[2]*Hunan Provincial Collaborative Innovation Center of High Power Fiber Laser, Changsha 410073, PR China*
*\*jmxu1988@163.com*
*\*\*zhoupu203@163.com*

**Abstract:** We present a hundred-watt-level linearly-polarized random fiber laser (RFL) pumped by incoherent broadband amplified spontaneous emission (ASE) source and prospect the power scaling potential theoretically. The RFL employs half-opened cavity structure which is composed by a section of 330 m polarization maintained (PM) passive fiber and two PM high reflectivity fiber Bragg gratings. The $2^{nd}$ order Stokes light centered at 1178 nm reaches the pump limited maximal power of 100.7 W with a full width at half-maximum linewidth of 2.58 nm and polarization extinction ratio of 23.5 dB. The corresponding ultimate quantum efficiency of pump to $2^{nd}$ order Stokes light is 89.01%. To the best of our knowledge, this is the first demonstration of linearly-polarized high-order RFL with hundred-watt output power. Furthermore, the theoretical investigation indicates that 300 W-level linearly-polarized single-mode $1^{st}$ order Stokes light can be obtained from incoherently pumped RFL with 100 m PM passive fiber.

## 1. Introduction

Random lasers are now attracting more and more attention and finding wide applications in many fields, such as illumination source, sensing technology and spectroscopic monitoring, due to its typical features of mode-free, structural simplicity and low coherence length [1-3]. In 2010, random fiber laser (RFL) based on the extremely weak Rayleigh scattering in a section of passive fiber to provide random distributed feedback (RDFB) was demonstrated for the first time [2]. In recent year, intensive investigations of RFLs have been reported on power scaling [4-7], linewidth narrowing [8-10], spectrum tuning [11-13], pulse operating [14-17], wavelength exploring [18, 19] and so on [3]. The novel application avenues of RFLs are also attempted in wide areas, such as frequency doubling [20], mid-infrared laser generation [21], and supercontinuum generation [22].

Especially, linearly-polarized laser is crucial for some practical application fields such as sensing and communication. And the linearly-polarized random lasing of RFLs have been widely investigated [16, 23-28]. Furthermore, the demonstration of high power RFL with linearly-polarized output was also attempted. Du et al. obtained linearly-polarized random lasing via linearly-polarized passive fiber and non-linearly-polarized pump source by introducing a certain fiber coiling technique [25]. Zlobina et al. [26] and Babin et al. [27] presented single-order and high-order linearly-polarized RFL via linearly-polarized passive fiber and linearly-polarized pump source with maximal output power of about ten-watt level and polarization extinction ratio (PER) of >22 dB. Besides, most of the reported high power RFLs employed coherent fiber lasers as the pump source [4-7, 25-27]. As we know, the demonstration of a truly continuous-wave fiber laser is, however, not trivial as self-pulsing can occur in several kinds of fiber lasers under specific pumping and cavity conditions [28-31]. And the characteristic frequencies of self-pulsing in fiber lasers, usually in the range of kilohertz to hundreds of megahertz, can transfer to the output Stokes light of RFL [32]. Then, a high degree of temporal coherence can be induced into the output light [33]. By contrast, incoherent amplified spontaneous emission (ASE) source from rare-earth doped fiber has been proved to have special features, such as low coherence, high temporal stability, and wide wavelength coverage [34, 35]. Comparing with common incoherent pump source for RFL, such as laser diode [36], high power pump light with high brightness can be obtained from ASE source. In our previous works, broadband ASE source has been utilized as the pumping source of linearly-polarized high-order RFL [37] and ultra-stable high-power mid-infrared optical parametric oscillator [38].

In this manuscript, we present a linearly-polarized $2^{nd}$ order RFL pumped by incoherent ASE source with maximal output power of 100.7 W and FWHM linewidth of 2.58 nm. The ultimate quantum efficiency of pump to $2^{nd}$ order Stokes light reaches 89.01%. The PER of the output Stokes light can achieve as high as about 23.5 dB, despite of the relative low PER value of pump source (about 15 dB). The standard deviation (STD) and peak-vale (P-V) value of the $2^{nd}$ order Stokes light at maximal power is 0.7% and 2.92%, respectively. To the best of our knowledge, this is the first demonstration of linearly-polarized high-order RFL with hundred-watt output power. And the theoretical study indicates that 300 W-level linearly-polarized single-mode $1^{st}$ order Stokes light can be obtained from the incoherently pumped RFL with 100 m polarization-maintained (PM) passive fiber.

## 2. Experimental setup and results

The experimental setup of the linearly-polarized high-order RFL system is depicted in Fig. 1. Pump source we employed is a high power linearly-polarized broadband ASE source in all-fiberized master oscillator power amplifier (MOPA) structure, which is similar with our previous reported linearly-polarized broadband ASE source in Ref [39]. The difference is that the three-stage power amplifiers simultaneously employed single-mode fiber exhibiting core diameter of 10 μm and numerical aperture (NA) of 0.08. Maximal output power of 127.7 W pumping ASE light can be obtained from this linearly-polarized broadband ASE source. The central wavelength and FWHM linewidth is 1074.8 nm and 9.1 nm, respectively. The PER of the ASE source at different power level maintains well with about 15 dB. As to the linearly-polarized high-order RFL, half-opened cavity structure, which is composed by two narrowband high reflectivity (99%) PM fiber Bragg gratings (FBGs) and a piece of 330 m long PM passive fiber, is utilized to decrease the threshold of random lasing [11]. The FBGs are utilized to select the radiation section of initial spontaneous Raman noise to obtain narrowband output. The central wavelength and FWHM linewidth of FBG 1 is 1120 nm and 1.5 nm. The corresponding central wavelength and FWHM linewidth of FBG 2 is 1178 nm and 2.16 nm. The passive fiber is employed to provide RDFB and Raman gain for the random lasing. The pigtailed fiber of FBGs and passive fiber are the same type of pigtailed fiber of

ASE source with core diameter of 10 μm and NA of 0.08. For appropriate heat management, the passive fiber is evenly winded on the outside surface of an air-cooled metal cylinder.

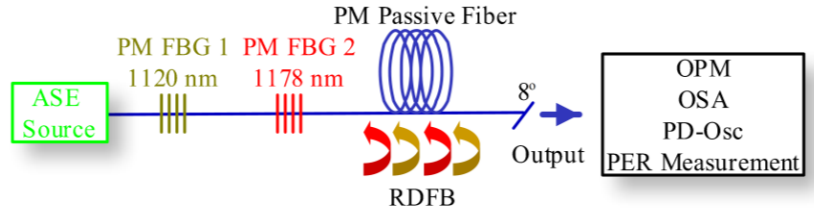

Fig. 1. Schematic of the linearly-polarized high-order RFL system. ASE, amplified spontaneous emission; PM FBG, polarization maintained fiber Bragg grating; RDFB, random distributed feedback; OPM, optical power meter; OSA, optical spectrum analyzer; PD, photon detector; Osc, oscilloscope; PER, polarization extinction ratio.

Figure 2 plots the output power and spectral characteristics of the linearly-polarized high-order RFL. Below the threshold of $1^{st}$ order Stokes (36.3 W), only transmitted pump light can be observed in the output light, and the attenuation ratio of FBGs and passive fiber for the broadband pump light can be calculated to be 12.56%. With the scaling of pump power from 36.3 W, the $1^{st}$ order Stokes power grows rapidly with the pump power up to the $2^{nd}$ order Stokes threshold (71.6 W) and then starts to decrease. For the multi-cascaded Rayleigh scattering (RS) - stimulated Brillouin scattering (SBS) generation [2, 40], broadband emission spectrum can be observed at the threshold of $1^{st}$ order Stokes, as shown in Fig. 2(b). The spectrum of the $1^{st}$ order Stokes light evolves to smooth and narrow with the scaling of pump power to 51.8 W. The output power of $1^{st}$ order Stokes light reaches the maximal value of 56.92 W with 70.3 W pump power injected and 3.09 W pump light remained. The corresponding optical-to-optical conversion efficiency and quantum efficiency of pump light to $1^{st}$ order Stokes light is 84.69% and 88.25%, respectively. The oscillation of broadband $2^{nd}$ order Stokes light can be measured at 71.6 W pump level. The broadband output spectrum evolves to smooth and narrowband shape when the pump power reaches 80.4 W. The maximal output power of the linearly-polarized high-order RFL is 104.8 W with 127.7 W pump light employed, which is limited by the available pump power. At maximal pump level, almost all of the pump light conversed to the high order Stokes light, and the residual pump light and $1^{st}$ order Stokes light is lower than the $2^{nd}$ order Stokes light with 21 dB and 26 dB, respectively. By numerical integrating, the $2^{nd}$ order stokes, $1^{st}$ order Stokes and residual pump component is calculated to be 100.7 W, 0.4 W, and 3.7 W. The optical-to-optical conversion efficiency of pump light to high-order Stokes light is 81.21%. And the corresponding quantum efficiency is 89.01%. Comparing with the results of reported ultimate efficiency linearly-polarized high-order RFL [27, 37], the quantum efficiency in this presented linearly-polarized high-order RFL pumped by broadband ASE source sets record value for the $2^{nd}$ order Stokes light, which is only slightly lower than the maximum quantum efficiency demonstrated for the $1^{st}$ order Stokes light (about 92.90 %) [7]. The relative low quantum efficiencies of the reported high order RFLs may be influenced by the relative high cavity loss. And the corresponding cavity loss of the high order RFL demonstrated in Ref. [27] and [37] is about 15% and 13%, respectively. And the following simulation will also support this opinion.

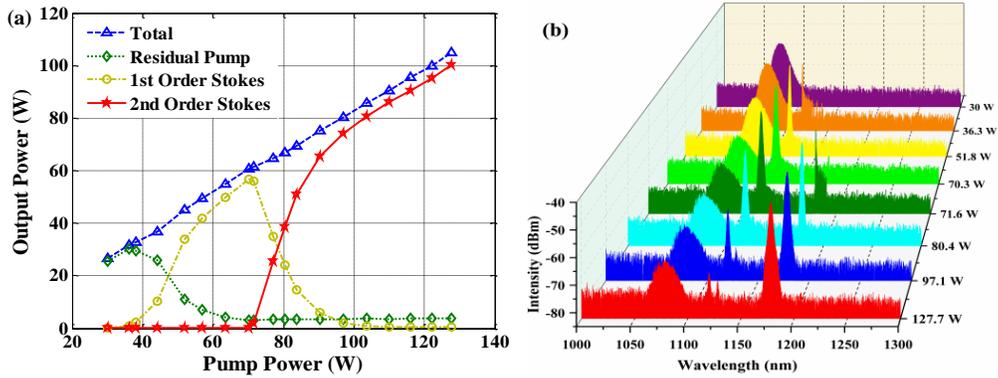

Fig. 2. (a) Output power as a function of pump power. (b) Output spectra at different power level.

The evolution of FWHM linewidth and polarization characteristics of output Stokes light as function of pump power is plotted in Fig. 3. For the 1st order Stokes light, the FWHM linewidth maintains well with ~1.4 nm, which is slightly narrower than the bandwidth of FBG 1. Additionally, no obvious linewidth broadening can be observed for the relative short fiber employed in the scheme and the relative low operating power level for the lasing of 1st order Stokes. As to the 2nd order Stokes light, obvious linewidth broadening can be measured with the scaling of operating power from 80.4 W to 103.6 W, and the FWHM linewidth evolves to a stable value of about 2.5 nm with the enhancement of pump power to 127.7 W. The linewidth broadening may be induced by the nonlinear effects such as self-phase modulation (SPM) and cross-phase modulation (XPM) [27].

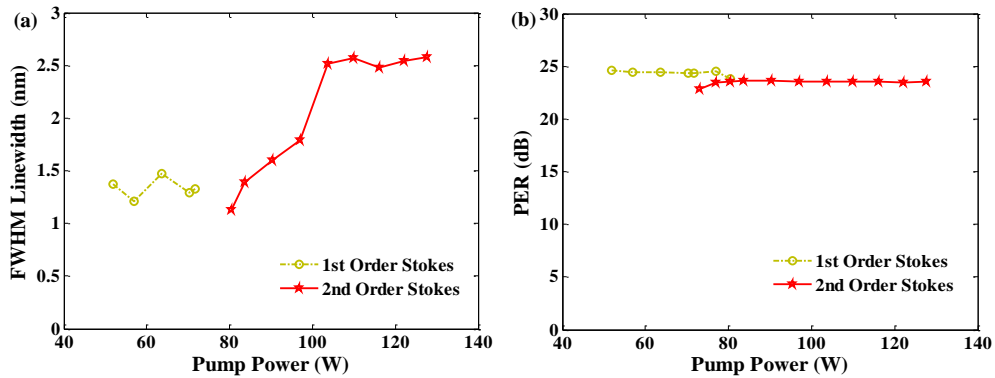

Fig. 3. (a) FWHM linewidth as a function of pump power. (b) PER evolution in the power scaling process.

The PER of the 1st and 2nd order Stokes light is measured by spatial PER measurement. After the collimation of output light by a spatial collimator, high reflection mirror operated at 1070 nm with a bandwidth of ± 10 nm and dichroic mirror for 1120 nm/1178 nm wavelength are employed to pick out the residual pump light from the Stokes light and split the high order Stokes light. Half-wave plates and polarization beam splitters operating at corresponding wavelength are utilized after the spectrum section splitting. The power along the two polarization axis are measured by two optical power meters (OPM) and the PER value can be calculated. Although the PER value of the broadband pump source is only about 15 dB, the PER of the output 1st and 2nd order Stokes light can achieve as high as about 23.5 dB. The performance improvement on the linearly-polarized characteristic of the Stokes light maybe induced by the random lasing as the Raman gain is polarization sensitive [41]. Furthermore, the linearly-polarized characteristics of the 1st and 2nd order Stokes light maintain well in the power scaling process without obvious fluctuation.

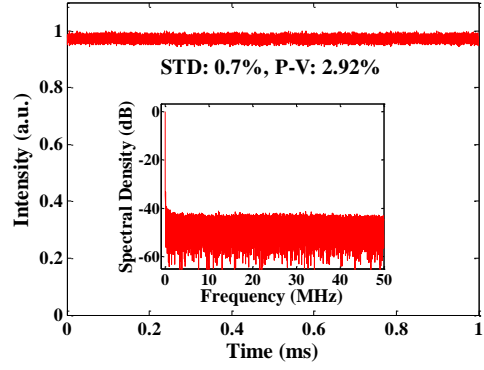

Fig. 4. The temporal signals and corresponding spectral densities of the RFL at maximal pump power.

Figure 4 shows the characteristics of high-order Stokes light in temporal and frequency domain at maximum power level. The temporal properties is measured by InGaAs photo-detectors (5 GHz bandwidth, rise time <70 ps) and digital phosphor oscilloscope (1 GHz bandwidth, 5 GS/sec sampling rate) with the help of spectrum splitting mirrors and fiber tapper. Thanks to the good stability of broadband ASE source [34, 35], stable linearly-polarized high-order Stokes light can be obtained. The STD and PV value of the temporal measurements at maximal pump level is 0.7% and 2.92%, respectively. Spectrum analysis (enclosed inside Fig. 4) shows that there no typical frequency content, which also shown the superior performance of this broadband ASE source pumped RFL in power stability.

## 3. Theoretical prospect of power scaling potential

The operation of this linearly-polarized RFL pumped by broadband incoherent ASE source is also investigated theoretically based on the classical steady-state light propagation equations [5, 7].

$$\pm \frac{dP_0^\pm}{dz} = -\frac{\lambda_1}{\lambda_0} g_{R1}(P_1^+ + P_1^- + 4h\upsilon_1 \Delta\upsilon_1 B_1)P_0^\pm - \alpha_0 P_0^\pm + \varepsilon_0 P_0^\mp \quad (1)$$

$$\pm \frac{dP_1^\pm}{dz} = g_{R1}(P_0^+ + P_0^-)(P_1^\pm + 2h\upsilon_1 \Delta\upsilon_1 B_1) + \varepsilon_1 P_1^\mp$$
$$- \frac{\lambda_2}{\lambda_1} g_{R2}(P_2^+ + P_2^- + 4h\upsilon_2 \Delta\upsilon_2 B_2)P_1^\pm - \alpha_1 P_1^\pm \quad (2)$$

$$\pm \frac{dP_2^\pm}{dz} = g_{R2}(P_1^+ + P_1^-)(P_2^\pm + 2h\upsilon_2 \Delta\upsilon_2 B_2) + \varepsilon_2 P_2^\mp - \alpha_2 P_2^\pm \quad (3)$$

$$B_j = 1 + \frac{1}{\exp\left[\frac{h(\upsilon_j - \upsilon_{j-1})}{k_B T}\right] - 1} \quad (j=1,\ 2) \quad (4)$$

Where $P$ denotes the power in different position and the subscript 0, 1, 2 stands for the pump light, the 1st order Stokes light, and the 2nd order Stokes light, respectively. Superscript + and - represents the forward and backward propagating waves, correspondingly. $g_R$, $\alpha$ and $\varepsilon$ is the Raman gain coefficient, signal loss and Rayleigh backscattering coefficient, respectively. $h$ is Planck constant; $\upsilon$ is the wave frequency; $\Delta\upsilon$ is the bandwidth of Stokes light. The parameter $B$ represents the population of the photon that introduces the noise from spontaneous Raman scattering.

**Table 1. Parameters for the Numerical Calculations**

| Parameter | Value | Unit |
|---|---|---|
| $\lambda_0, \lambda_1, \lambda_2$ | 1074.8, 1120, 1178 | nm |
| $g_{R1}, g_{R2}$ | 0.533, 0.792 | km$^{-1}$W$^{-1}$ |
| $\alpha_0, \alpha_1, \alpha_2$ | 3.8, 3.3, 2.8×10$^{-4}$ | m$^{-1}$ |
| $\varepsilon_0, \varepsilon_1, \varepsilon_2$ | 6.8, 6, 5.2×10$^{-7}$ | m$^{-1}$ |
| $\Delta\upsilon_1, \Delta\upsilon_2$ | 0.22 | THz |
| T | 298 | K |
| $R_{L1}, R_{L2}$ | 0.99 | |
| $R_{R1}, R_{R2}$ | 2.6×10$^{-6}$ | |

The boundary conditions can be described as $P_0(0)=P_{in}$, $P^+_{1,2}(0)= R_{L1,2}P^-_{1,2}(0)$, $P^-_{1,2}(L)= R_{R1,2}P^+_{1,2}(L)$, where $P_{in}$ is the input pump power, and $R_{L1,2}$ and $R_{R1,2}$ stands for the reflectivity at the left and right end, respectively. As the 3$^{rd}$ order Stokes light have not been observed in the experiment, the generation of 3$^{rd}$ order Stokes light is not considered in the numerical simulation. The values of parameters set in the numerical simulation are listed in Table 1.

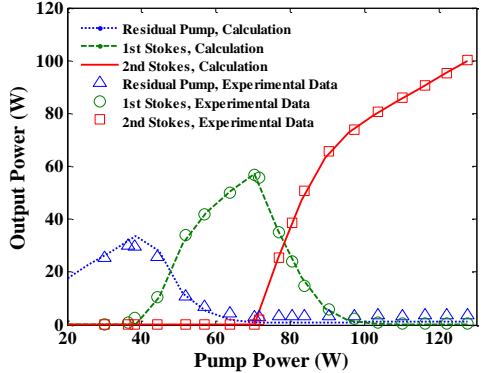

Fig. 5. Experimental measured and numerical calculated Stokes power of the incoherently pumped linearly-polarized RFL.

As depicted in Fig. 5, the results calculated by the model and parameters fit well with the experimental data, indicating that the presented model and parameters can be utilized to investigate this incoherently pumped linearly-polarized single-mode RFL. It should be noted that the value of coefficient $g_{R1}$ is lower than that of $g_{R2}$, which is different from previous results in Refs [5] and [7]. This may be explained as follows: the Raman gain peak of 1074.8 nm pump light corresponds to about 1128 nm, and the mismatching of Raman gain peak of 1120 nm 1$^{st}$ order Stokes light will experience a relative low Raman gain.

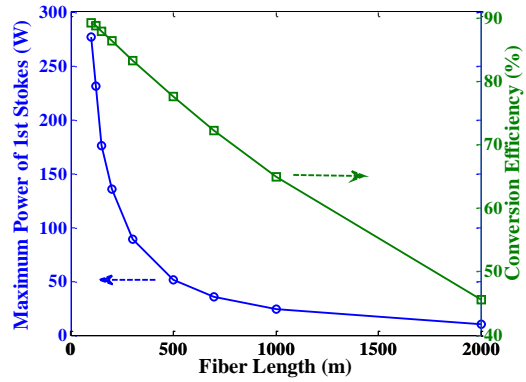

Fig. 6. Maximum output power of 1st order Stokes light and corresponding optical-to-optical conversion efficiency of the incoherently pumped linearly-polarized single-mode RFL as a function of the PM passive fiber length.

To investigate the power scaling ability of linearly-polarized RFL pumped by incoherent broadband ASE source, the maximum output power and corresponding optical-to-optical conversion efficiency of $1^{st}$ order Stokes light is calculated. Additionally, the maximal output power is defined as the highest power of the $1^{st}$ order Stokes light when further power enhancement is limited by the $2^{nd}$ order threshold. And the threshold of $2^{nd}$ order Stokes light refers to the pump power in the condition that the output power of $2^{nd}$ order Stokes light is 1% of the input power. It should be noted that the power boosting potential of the $2^{nd}$ order Stokes light is not investigated for the lacking of the parameters of $3^{rd}$ order Stokes light in this single-mode PM passive fiber presently. To suppress the generation of $2^{nd}$ order Stokes light and obtain more $1^{st}$ order Stokes light, the high reflectivity FBG corresponding to the $2^{nd}$ order Stokes light is removed from the left side and the parameter $R_{L2}$ in the simulation is set to be $2.6 \times 10^{-6}$, which is the same with the value of $R_{R2}$. The simulation results are displayed in Fig. 6. With the decreasing of fiber length, the maximal output power of $1^{st}$ order Stokes light increases exponentially, which is similar with the calculation results reported in Ref [5] for random-polarized RFL constructed by G. 652 passive fiber. What's more, the corresponding optical-to-optical conversion efficiency enhances with the decreasing of fiber length for the lower attenuation of pump and Stokes light. With 100 m PM single-mode passive fiber (the same with the length limitation of passive fiber in Ref [5]) and 313 W linearly-polarized incoherent pump light utilized, 278 W maximum output power of linearly-polarized single-mode $1^{st}$ order random laser can be obtained.

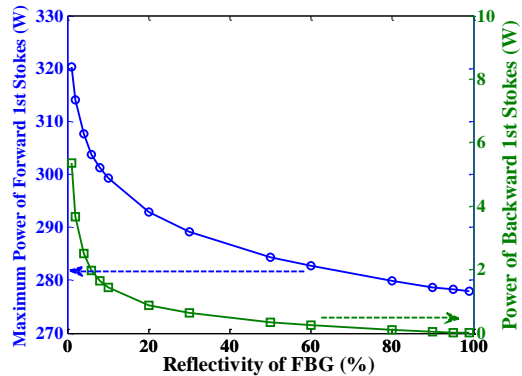

Fig. 7. Maximum output power of forward $1^{st}$ order Stokes light and corresponding backward $1^{st}$ order Stokes power of the incoherently pumped linearly-polarized single-mode RFL as a function of the reflectivity of FBG.

As the reflectivity plays an important role on the random lasing [42], further power scaling of this incoherently pumped linearly-polarized RFL with 100 m PM single-mode passive fiber can also be realized via the optimization of the reflectivity of the FBG ($R_{L1}$) employed for the constructing of half-opened cavity. It should be noted that the decreasing of $R_{L1}$ would introduce the leaking of backward 1$^{st}$ order Stokes light. The maximum output power of forward 1$^{st}$ order Stokes light from the angle-cleaved-side of the passive fiber and corresponding backward 1$^{st}$ order Stokes light leaking from the FBG as a function of the reflectivity of FBG is depicted in Fig. 7. Fortunately, the maximum power of forward 1$^{st}$ order Stokes light increases with lower reflectivity of FBG; unfortunately, the backward 1$^{st}$ order Stokes light enhances drastically simultaneously, which may lead to the damaging of high power pump source. By employing high reflectivity (99%) FBG, the maximum power of forward 1$^{st}$ order Stokes light and corresponding power of backward 1$^{st}$ order Stokes light is 278 W and neglectable 5 mW; with the decreasing of reflectivity of FBG to 1%, the maximum forward 1$^{st}$ order Stokes power and corresponding backward 1$^{st}$ order Stokes power is 320 W and 5.4 W, respectively. Therefore, the backward power isolation of pump source is necessary for high power RFL in half-opened cavity with a low-reflectivity FBG. Generally speaking, 300 W-level incoherently pumped linearly-polarized single-mode RFL can be realized with properly backward light isolation.

## 4. Summary

In summary, we have presented an incoherently pumped linearly-polarized high-order RFL with hundred-watt output power. The pump source we employed is a linearly-polarized broadband ASE source with maximal output power of 127.7 W. The central wavelength and FWHM linewidth of the pump light is 1074.8 nm and 9.1 nm, respectively. The linearly-polarized high-order RFL employs half-opened cavity structure, which is composed by two high reflectivity FBGs and a section of PM passive fiber with a length of 330 m. The maximal output power of the 2$^{nd}$ order Stokes light is 100.7 W, which is limited by the available pump power. The quantum efficiency of pump to 2$^{nd}$ order Stokes light reaches as high as 89.01%, which is the highest value ever reported for high-order RFLs. The FWHM linewidth of the 2$^{nd}$ order Stokes light at maximal power level is 2.58 nm. The PER value of the Stokes light can reach as high as about 23.5 dB, despite of the relative low PER value of pump source (about 15 dB). Thanks to the good stability of pumping broadband ASE source, the STD and P-V value of the high order Stokes light at maximal power level is 0.7% and 2.92%, respectively. The demonstration indicates that high-power narrowband random laser can be obtained efficiently by the pumping of broadband ASE source. And the theoretical simulation evaluates that 300 W-level linearly-polarized single-mode 1$^{st}$ order random laser can be obtained with 100 m PM single-mode passive fiber. To the best of our knowledge, this is the first demonstration of linearly-polarized high-order RFL with hundred-watt-level output. And this linearly-polarized RFL with good temporal and polarization performance may find special advantages in industrial processing and mid-infrared laser generations as well. Further power scaling is available with more powerful pumping ASE source and optimization of operation parameters.

**Funding**

National Natural Science Foundation of China (NSFC) (61322505, and 61635005).

**Acknowledgments**

We are particularly grateful to Xueyuan Du, Xiaolin Wang, Rongtao Su, Xiaoxi Jin and Wei Liu for their supports on this work.